\documentclass[mathleft,usenatbib,useAMS
]{an}
\usepackage{graphicx}
\usepackage{times}
\usepackage{color}
\usepackage{amsmath,amssymb}

\def\sdss{{SDSS~J0959+1259}}
\def\J1{{SDSS~J1038+3921}}
\def\J2{{SDSS~J1626+1422}}
\def\xmm{{\it XMM-Newton\/}}

\begin{document}

\Yearpublication{2016}%
\Yearsubmission{2016}%

\title{Unveiling multiple AGN activity in galaxy mergers}

\author{A. De Rosa\inst{1}\fnmsep\thanks{Corresponding author:
  \email{alessandra.derosa@iaps.inaf.it\newline
  \texttt{www.issibern.ch/teams/agnactivity/Home.html}}\newline}
S. Bianchi\inst{2}, T. Bogdanovi\'c\inst{3},  R. Decarli\inst{4}, J. Heidt \inst{5}
R. Herrero-Illana\inst{6,7}, B. Husemann\inst{8,9}, S. Komossa\inst{10}, 
E. Kun\inst{11}, 
N. Loiseau\inst{12}, 
M. Guainazzi\inst{13}
Z. Paragi\inst{14}, 
M. Perez-Torres\inst{6,15}, 
E. Piconcelli\inst{16}, 
K. Schawinski\inst{17},  
C. Vignali\inst{18}
}
\titlerunning{Multiple AGN Activity}
\authorrunning{A. De Rosa}
\institute{
INAF/Istituto di Astrofisica e Planetologie Spaziali. Via Fosso del Cavaliere - 00133 Roma - Italy
\and 
Dipartimento di Matematica e Fisica, Universit\`a degli Studi Roma Tre, via della Vasca Navale 84, 00146 Roma, Italy
\and
Center for Relativistic Astrophysics, Georgia Institute of Technology, Atlanta, GA 30332, USA
\and
Max-Planck Institut fuer Astronomie, Germany
\and
ZAH Landessternwarte Königstuhl. D-69117 Heidelberg. Germany 
\and
Instituto de Astrof\'{\i}sica de Andaluc\'{\i}a (IAA-CSIC), 18008 Granada, Spain
\and
European Southern Observatory (ESO), Alonso de Cordova 3107, Vitacura, Casilla 19001, Santiago de Chile, Chile
\and
 MPIA, Konigstuhl 17 D-69117 Heidelberg, Germany
\and
European Southern Observatory Headquaters, Karl-Schwarzschild-Str. 2, 85748 Garching, Germany
\and
QianNan Normal University for Nationalities, Longshan Street, Duyun City of Guizhou Province, China
\and
 Department of Experimental Physics, University of Szeged, D\'om t\'er 9, H-6720 Szeged, Hungary
\and
ESAC/XMM-Newton Science Operations Centre, Spain
\and
ESA - European Space Agency. ESTEC, Keplerlaan 1 2201AZ Noordwijk, The Netherlands
\and
Joint Institute for VLBI ERIC, Postbus 2, NL-7900 AA Dwingeloo, The Netherlands
\and
Visiting Scientist at the Depto. de F\'isica Te\'orica, Facultad de Ciencias, Univ. de Zaragoza, Spain
\and
Osservatorio Astronomico di Roma (INAF), via Frascati 33, 00040 Monte Porzio Catone ( Roma), Italy 
\and
Institute for Astronomy, Wolfgang-Pauli-Str. 27, 8093 Zürich, Switzerland
\and
Dipartimento di Fisica e Astronomia, Universit\`a degli studi di Bologna, Viale Berti Pichat
6/2, 40127, Bologna, Italy
}

\received{4 October 2016}
\accepted{4 October 2016} 


\abstract{%
 In this paper we present an overview of the MAGNA (Multiple AGN Activity) project aiming at a comprehensive study  of multiple supemassive black hole systems. With the main goal to characterize the sources in merging systems at different stages of evolution, we selected a sample of objects optically classified as multiple systems on the basis of  emission line diagnostics and  started a massive multiband observational campaign. Here we report on the discovery of the exceptionally high AGN density compact group \sdss. A multiband study suggests that strong interactions are  taking place among its galaxies through tidal forces, therefore  this system represents a case study for physical mechanisms that trigger nuclear activity and star formation. We also present a preliminary analysis of the multiple AGN system SDSS~J1038+3921.}

\maketitle

\begin{table*}[h!]
\centering
\caption{\footnotesize{The sources of the multiple systems analysed in this paper: \sdss\ and SDSS J1038+3921.}}
\begin{tabular}{cccccccc}
\hline
 & source & angular sep.  & proj. distance & z  & Lx &  L(OIII) & type \\
 & SDSS & (1) & (2) & (3) & (4) & (5) & (6)   \\
\hline
I-src1 & J095906.68+130135.4 & 327  & 227 &    0.037 &   5 &     0.07 &   NL AGN \\
I-src2 & J095903.28+130220.9 & 377 & 262 &  0.036  &  0.12 &     0.003 &  LINER  \\
I-src3 & J095908.95+130352.4 &  318 & 221 &  0.0339 &   0.1   &      0.0008 & LINER\\
I-src4 & J095912.19+130410.5 &  280 & 195 & 0.0337 & $<$0.3&    0.002 &   SFG \\
I-src5 & J095914.76+125916.3 &  259 &180 &  0.034 &   2.8 &  0.4 &    NL AGN \\
I-src6 & J095859.91+130308.4  &  430 & 299 &  0.035 &  $<$0.2 &   0.008 &  SFG\\
I-src7 & J095900.42+130241.6   & 418 & 290 &  0.035 &   $<$0.1 &     0.02 &  SFG \\
I-src8 &  J095955.84+130237.7 &  389 & 270 &  0.035 &  15 & 0.03 & BL AGN \\
\hline
II-src1 & J103853.29+392151.2&   40.4 & 42.9& 0.0548 &   3.4 &    0.04 &  BL AGN \\
II-src2 & J103855.95+392157.6&      	-	&   -	& 		-		&   4$^\star$  	&     0.1 & NL AGN   \\
\hline
\end{tabular}
\label{tab:fluxes}

\footnotesize{{{${(1)}$Angular separation between pairs in arcsec. In the case of \sdss\ it is measured with respect to the center of the system (149.87$^\circ$;13.03$^\circ$) ${(2)}$Projected distance in kpc. ${(3)}$Redshift. ${(4)}$ 2-10 keV unabsorbed luminosity  in 10$^{42}$ erg s$^{-1}$. $^\star$Assuming a Compton thick spectrum with N$_H$=10$^{24}$ cm$^{-2}$. ${(5)}$ De-reddened  [OIII] luminosity in 10$^{42}$ erg s$^{-1}$. $^{(6)}$ Object Type. }}}
\end{table*}

\section{Introduction}
Hierarchical merger models of galaxy formation predict that binary AGN should be common in galaxies (Haehnelt et al. 2002; Volonteri et al. 2003).  
Understanding the types of galaxies and specific merger stages where  AGN pairs occur provides important clues about the peak of the black hole growth during the merging process (Begelman et al.1980; Escala et al. 2004).  
First compact AGN pairs in advanced mergers and interacting galaxies have been identified in 
X-rays (e.g., Komossa et al. 2003, Bianchi et al. 2008, Piconcelli et al. 2010, Guainazzi et al. 2005) and in the radio
band (Rodriguez et al. 2006). Given the small number of known systems, there is a need
of increasing the sample size significantly, in order to cover a wide range of  spatial separations
and stages of galaxy mergers. Therefore, the search for AGN pairs has recently received great attention
(see Bogdanovic 2015 and Komossa \& Zensus 2016 for recent reviews), and different
methods have been proposed to identify good candidates. With the availability of massive spectroscopic
surveys like the Sloan Digital Sky Survey (SDSS), candidate AGN pairs have been identified from optical double-peaked
emitters. In these systems, the narrow line emission is double-peaked, and one possible interpretation is that they  consist of a pair of AGN.
Multi-wavelength follow-up campaigns have shown that only a small fraction of these systems actually harbour AGN
pairs (e.g., Fu et al. 2011, Fu et al. 2012, Smith et al. 2012, Comerford et al. 2012), while the
majority of the double-peakers is produced by other mechanisms like two-sided outflows,
or jet-cloud interactions. Follow-up observations of candidate pairs in the infrared, optical, radio and X-rays are therefore
essential in finding the true pairs.\newline
The key goal of our MAGNA (Multiple AGN Activity) project is to determine the multiband properties (radio,  IR, optical and X-rays) of a homogeneously selected sample of AGN systems in different stages of merging. 
As a first step, we identified the systems from the sample of AGN pairs optically selected from the SDSS-DR7 by Liu et al. (2011).
We selected the optically classified Seyfert-Seyfert pairs using the  [OIII]/H$\beta$ line-ratio diagnostic (Baldwin et al. 1981). Finally, we chose  the maximum projected separation of a pair of sources to be $<$ 60 kpc to remove from the sample  many non-interacting AGN pairs (Satyapal et al. 2014). This separation threshold allows us to keep a significant number of wide pairs. The final  ``master sample" consists of 16 systems, all of them harbouring at least two  AGN (Sy-Sy type, following the optical classification) with redshift between 0.03-0.17. 
We started  intensive observational campaigns (VLA, EVN, MUSE, XMM) on the master sample.

In this paper we report on the study of multiple SMBH systems investigated in our MAGNA programme.
Data analysis is presented in Sect. \ref{sect:data}. In Sect. \ref{sect:j0959}  we report on the discovery of the exceptionally high AGN density compact group \sdss\ while a preliminary analysis of the multiband data of the system SDSS~J1038+3921 is presented in Sect. \ref{sect:j1038}. Our conclusions and future prospects are described in Sect. \ref{sect:future}.
We assume H$_{0}$ = 70 km\,s$^{-1}$\,Mpc$^{-1}$, $\Omega_\Lambda$= 0.7, $\Omega_M$ = 0.3, and AB magnitudes. 
\begin{figure*}[t!]
\centering
\includegraphics[width=65mm,height=77mm]{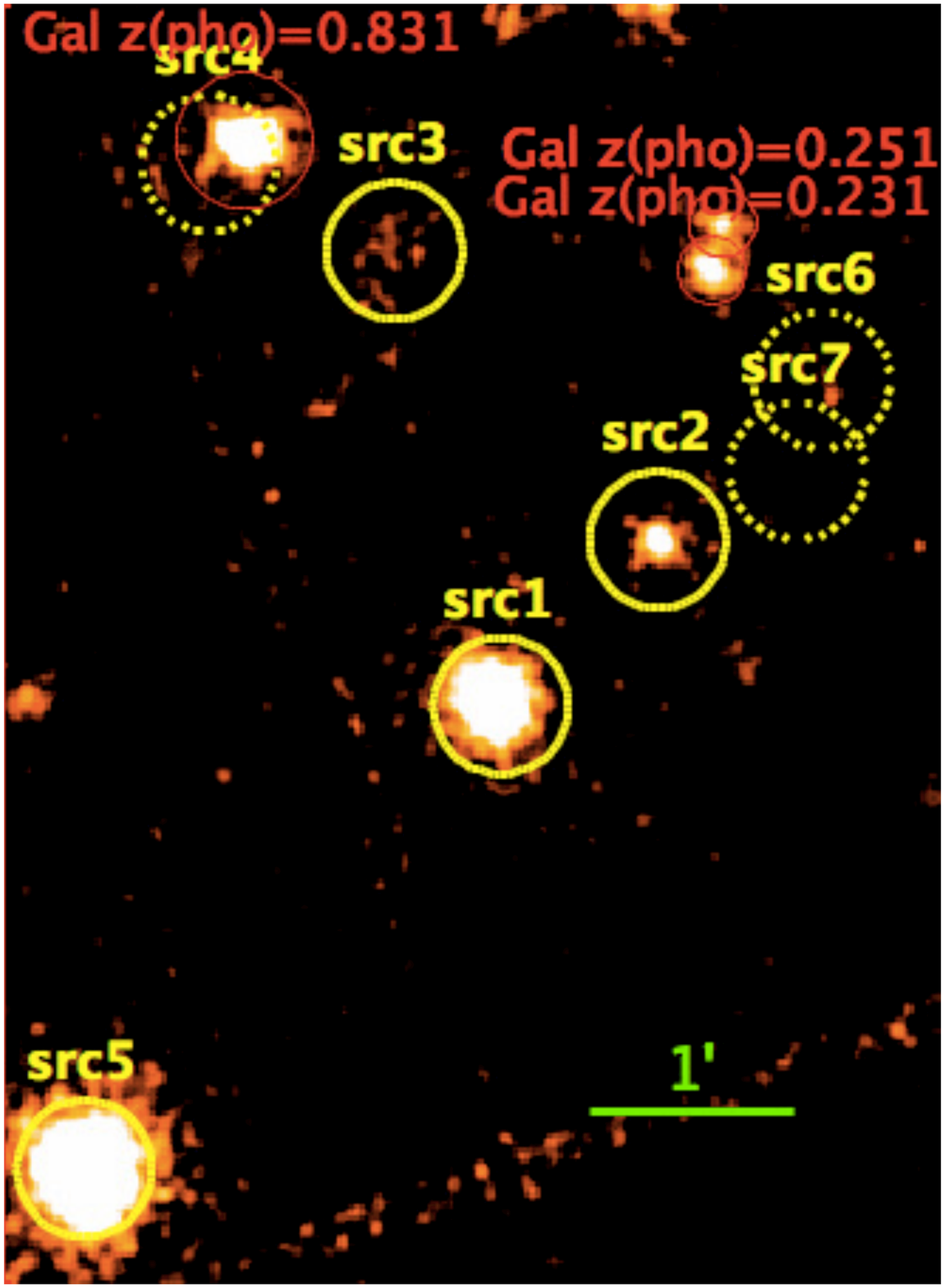}
\includegraphics[width=65mm,height=77mm]{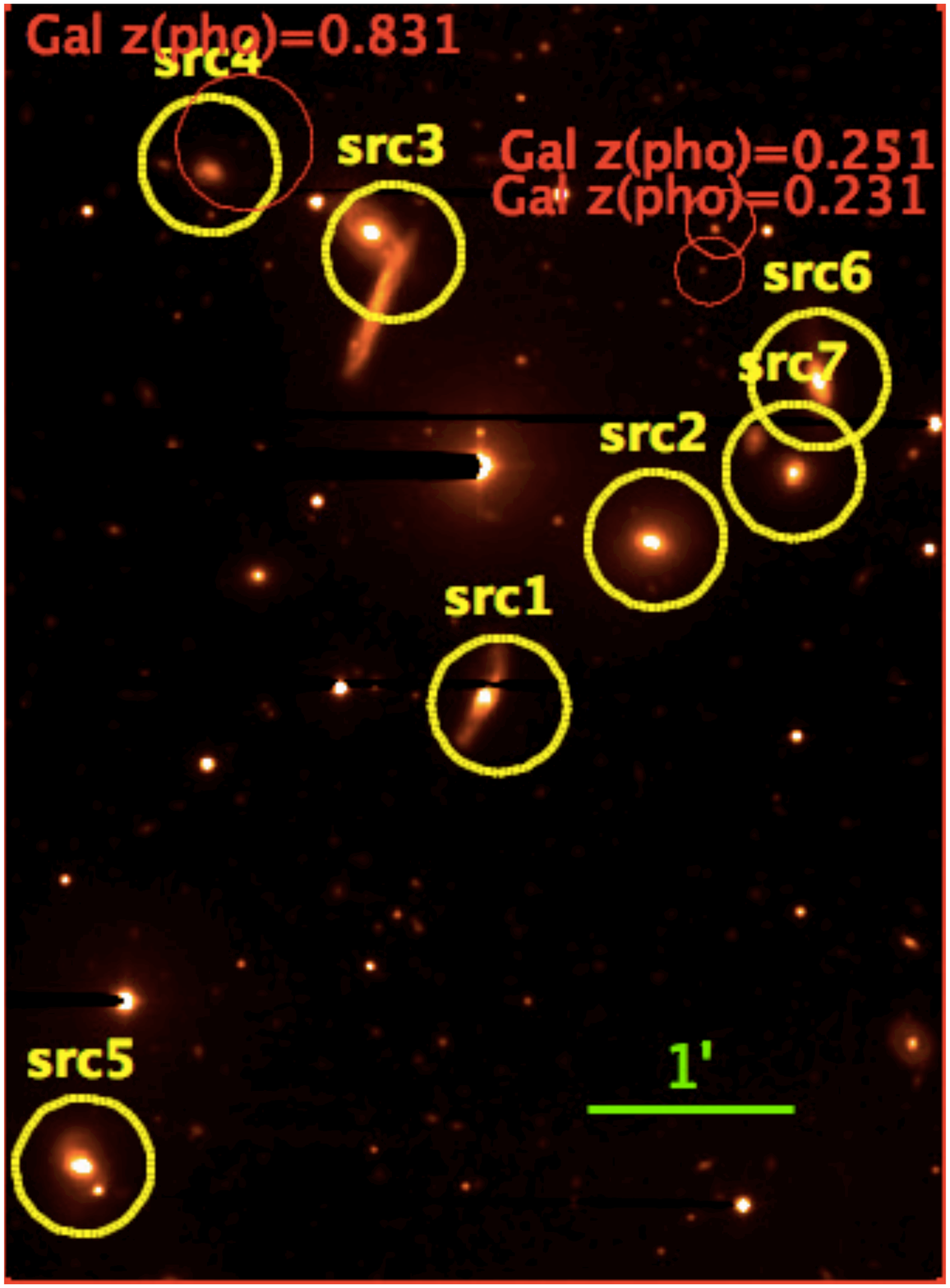}
{\footnotesize\caption{{\it Left:} XMM-EPIC smoothed mosaic image of the CG region (pn, MOS1 and MOS2 co-added). North is up and East to the left.  I-src8 is off the plot, 10 arcmin to the East from src1. Yellow circles show the group sources while undetected sources are shown with a dashed line. Red circles are the sources in the field not included in our group lacking a spectroscopic redshift in the SDSS (our estimated photometric redshifts based on BUSCA images suggest that they are background objects). {\it Right:} The  BUSCA R-band image of the same region on the left. Labels are the same.}}
\label{fig:ima0959}
 \end{figure*}
\section{MAGNA: Observations and data analysis}
\label{sect:data}

\subsection{X-rays}
The sources in the systems observed with \xmm\  were detected using the EPIC source finding threads \texttt{edetect chain}, on 5 images in the $0.3-0.5$ keV, $0.5-1$ keV, $1-2$ keV, $2-4.5$ keV, $4.5-12$ keV energy bands with a detection threshold of 3$\sigma$. 
All spectra were extracted from circular regions (with radius ranging from 15'' to 30'' depending on the source counts) which include  more than 80\% of the source counts at 1.5 keV in the EPIC cameras. The background spectra were extracted in the same CCD chip from circular regions free from contaminating objects and of the same size as the regions containing the source.
All events were screened  and filtered for the flaring events. Each spectrum (and associated background) was rebinned in order to have at least 25 counts for each background-subtracted spectral channel and not to oversample the intrinsic energy resolution by a factor larger than 3.
Spectral fits for pn and (co-added) MOS cameras were performed in the 0.3--10 keV energy band.

\subsection{Optical}
In order to identify the multiple sources in the systems we used 
the most common optical emission-line diagnostic diagrams (BPT, Baldwin et al. 1981).
We retrieved the SDSS-III DR12 spectra for all sources in our systems  (see Tab. \ref{tab:fluxes}) from the survey webpage\footnote{http://skyserver.sdss3.org/dr12}.
The emission line flux of all the primary diagnostic lines (H$\beta$, [OIII] $\lambda5007$, [OI] $\lambda6300$,
H$\alpha$, [NII] $\lambda6583$ and [SII] $\lambda\lambda6713,6732$) were measured on top of the stellar continuum
using the package \textsc{PyParadise}. \textsc{PyParadise} models the stellar continuum as a superposition of template stellar population spectra from the CB07 library \cite{bruzual03} after normalizing both the SDSS and the template spectra with a running mean over 100pix, interpolating regions with strong emission lines. A simple Gaussian kernel  is used to match the template spectra to the line-of-sight velocity distribution. The line fluxes are then inferred by fitting the Gaussian line profiles coupled in redshift and intrinsic rest-frame  velocity dispersion. Errors are obtained using a bootstrap approach where 100 realizations of the spectrum were generated based on the pixel errors with just 80\% of the template spectra and modelled again with the  same approach (at fixed stellar kinematics). 

In order to better characterize the sources in the crowded field of \sdss\, we observed the region using the multi–band camera BUSCA on the 2.2\,m telescope in Calar Alto observatory. These observations span the whole optical window from U to I over a $12'\times12'$ field of view. The throughput curves of the dichroic mirrors, convolved with the detector efficiency, roughly correspond to those of the SDSS u, r, i+z, and Johnson B filters. Five 12 min-long frames were collected in each band, dithered by a few arcsec in order to clean our final images from cosmic rays and bad pixels. Full details on data analysis of the system \sdss\ are reported in De Rosa et al. 2015.

\subsection{Radio}
We used radio 1.4 GHz data from two public data sets: (1) the NRAO VLA Sky Survey (NVSS; Condon et al. 1998), which have an angular resolution of $\sim$45 arcsec and a completeness limit of 2.5 mJy, and (2) the Faint Images of the Radio Sky at Twenty-Centimeters Survey (FIRST; Becker et al. 1994) with a resolution of $\sim$5 arcsec and a typical rms of 0.15 mJy.

\section{The Compact Group SDSS~J0959+1259}
\label{sect:j0959}

The system associated with the \sdss\ at redshift z=0.035  was recognized as the only quintuplet detected in a sample of 1286 multiple AGN/LINER systems (\cite{liu11}). 
The galaxies in this field that constitute a compact group (CG) have projected separations of $\lesssim$100h$^{-1}$ kpc and the line-of-sight velocity differences of $\lesssim$500 km s$^{-1}$ (\cite{hickson97}). Within these criteria we find 7 spectroscopically identified sources (from I-src1 to I-src7 in Tab. \ref{tab:fluxes}) while another source in this region (I-src8 NGC~3080)  is close in redshift (z=0.035) but at larger projected separation, and it is included  in our study. Detailed study on \sdss\ is presented  in De Rosa et al. (2015).
We have analysed the multiwavelength properties of the system using radio, optical and X-ray data.
The mean redshift of the CG is $\langle$z$\rangle$=0.0353 and the mean velocity is $\langle$v$\rangle$=10593 km s$^{-1}$. The center of this field has RA 09h 59m 28.97s and DEC 13$^\circ$ 01' 53.0'', and the average distance of sources from the geometric center of the group is 240~kpc. 
The CG is shown in Fig. \ref{fig:ima0959}, where the composite EPIC-pn and MOS12 (left panel) and the R-band BUSCA images  (right panel) are presented. I-src8, which is located 10 arcmin to the East from src1, is not shown. Yellow circles indicate CG sources, the dashed yellow circles mark the three objects not detected in X-rays, while the red circles show the background sources in the field not related to the group. 
All three  BPT diagrams lead to a consistent classification of objects into two Seyfert 2s, two LINERs and three Star Forming Galaxies (SFGs). The optical spectrum of I-src8 shows broad Balmer emission lines (FWHM$\sim$750 km s$^{-1}$) and thus points to a type-1 Seyfert. 
\begin{figure}[t!]
\includegraphics[width=75mm,height=65mm]{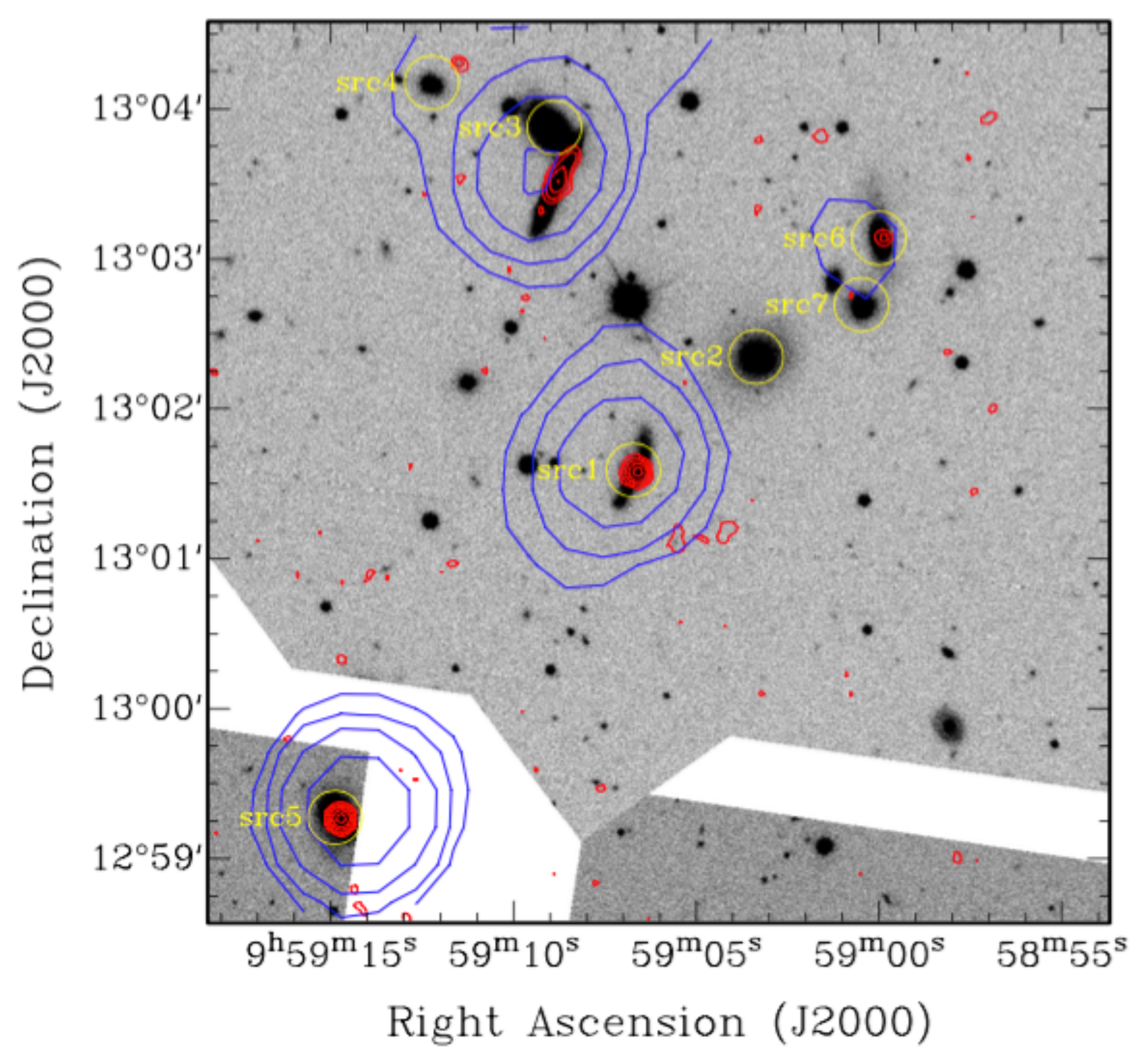}
{\footnotesize\caption{SDSS~J0959+1259. An overlay of the SDSS images of the compact group with the Very Large Array NVSS
(blue contours) and FIRST (red contours) survey maps.}}
\label{fig:nvss0959}
\end{figure}  
The 0.3$-$10 keV spectra of  the two Seyfert 2s  (I-src1 and I-src5) are modelled with an absorbed power-law plus a thermal component with temperature of kT$\sim$0.1-0.6 keV (\texttt{mekal}  in Xspec) emerging below 2 keV. The cold absorption gas has column densities in the range 
N$_{H}\approx$1-20 $\times$ 10$^{22}$ cm$^{-2}$. 
The broadband spectrum the Seyfert 1 (I-src8) is well fitted with an absorption component with partial covering fraction f$_{c}$=0.41$\pm$0.07. 
The range of luminosity of the Seyferts is 10$^{42}$--10$^{43}$ erg s$^{-1}$ and a narrow Fe K$\alpha$ emission line is also measured in their spectra, with an equivalent width of $\sim$100-130 eV.
The two LINERs (I-src2 and I-src3) are modelled with an unabsorbed power-law continuum in the X-ray band and show no detectable Fe line. Their 2--10 keV luminosity is nevertheless high ($\sim$10$^{41}$erg s$^{-1}$), above the mean value found in the systematic X-ray study of the largest sample of LINERs (\cite{gonzalez09}). This is a strong indication that LINERs in this CG may be accretion driven; in this case, the fraction of AGN in the CG rises from 40 to 60\% (from 3 to 5 out of 8). 
None of the SFGs (I-src4, I-src6, I-src7) are detected by XMM, placing a 3$\sigma$ upper limit on flux in 2-- 10\,keV of about 1.4$\times 10^{-14} {\rm erg\, cm^{-2} s^{-1}}$. This corresponds to a luminosity  of $\sim 4 \times10^{40} {\rm erg\, s^{-1}}$, assuming a photon index 1.7 and absorbing column density of 10$^{22}$ cm$^{-2}$.\
In order to complement the multiwavelength information, we  analyzed radio archival data, an overlay of SDSS optical as well as VLA NVSS and FIRST survey images is shown in Fig. 2.
The two Sy2 AGN (I-src 1 and I-src5) are both detected  at the 10--20 mJy level, and the FIRST/NVSS ratios
indicate these are quite compact on arcsec scales. The third well-established AGN (I-src 8) is indicating a resolved radio emission already on arcsec scales. The two LINERs are undetected on $\sim$mJy levels, and only one SFG is detected in FIRST 
but not in NVSS. A comparison of radio-to-X-ray luminosities indicates
that the two Sy2s are extremely radio-loud, and the Sy1 is radio-quiet, i.e. below logR$_{X}$=$-$4.5, with R$_{X}$=L$_R$/L$_X$ (Terashima \& Wilson 2003). 

The CG in \sdss\ represents one of the best examples of exceptionally strong nuclear activity in CGs in the nearby Universe. The only other example of an AGN  group rich like this is the well known HCG~16 (\cite{ribeiro96},\cite{turner01}), with an AGN fraction of 75\% (three out of four galaxies).
In addition, the clearly distorted shape of one member galaxy (I-src1) also hints on strong interactions within the group.
The high AGN fraction in the CG and the signature of interactions make therefore SDSS~J0959+1259 ideally suited to study nuclear activity and star formation triggered by tidal interactions in compact groups.
We then started a massive observational campaign on this CG, obtaining high resolution EVN + eMERLIN  and VLT-MUSE observations.
These combined observations will reveal further inside into the cause of the exceptionally strong
activity of this CG and what it implies for the evolution of its members (see details in Sect. \ref{sect:future}).
\begin{figure*}[t]
\centering
\includegraphics[width=85mm,height=60mm]{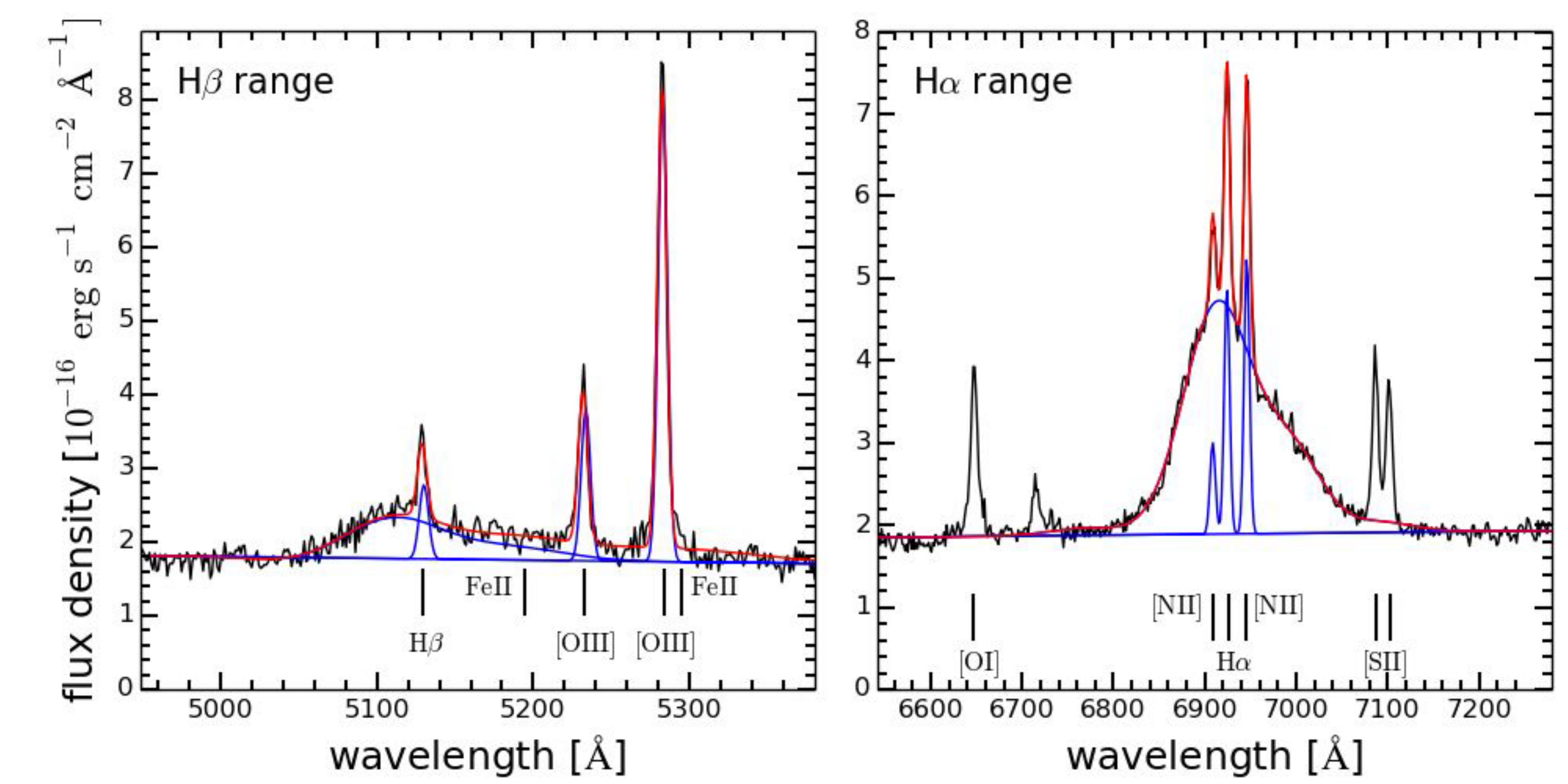}
\includegraphics[width=85mm,height=60mm]{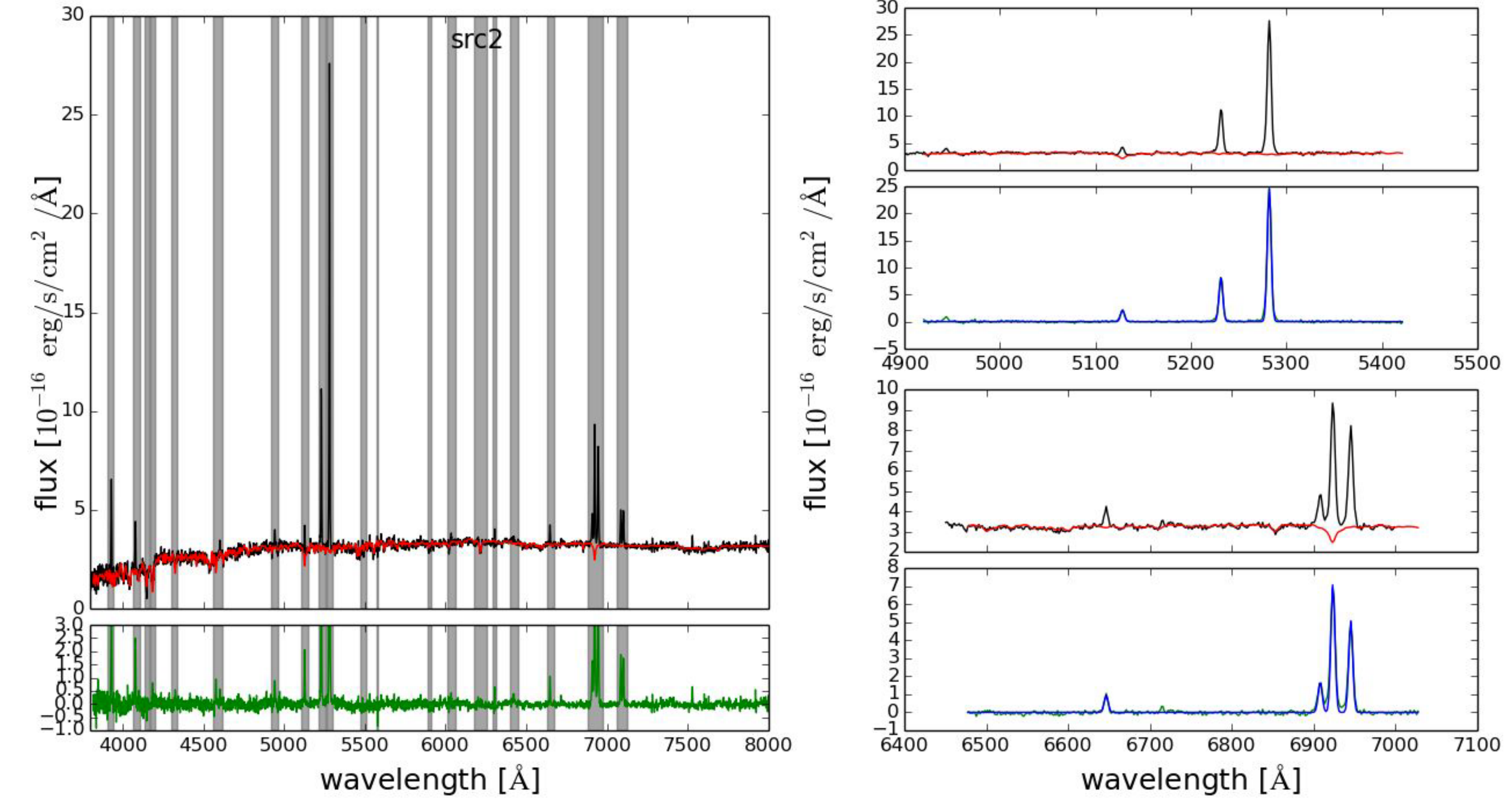}
{\footnotesize\caption{SDSS~J1038+3921. The SDSS spectra of the BL AGN (src1, left panels) and the NL AGN (src2, right panels). For the NL  we first model the stellar continuum with a super-position of single-stellar population templates and a simple Gaussian kernel for the line-of-sight kinematics. Then we fit the lines with single Gaussians coupled in the kinematics. For the BL  we  subtract a quasi-linear continuum from the adjacent continuum of the broad Balmer lines. Afterwards we fit a high-order Gauss-Hermite polynom to the broad component and a double Gaussian for the narrow lines but coupling all lines with the same kinematics.}}
\label{fig:1038_opt}
 \end{figure*} 
\section{The multiple SMBH system in SDSS J1038+3921}
\label{sect:j1038}

\begin{figure*}[t]
\centering
\includegraphics[width=80mm,height=65mm]{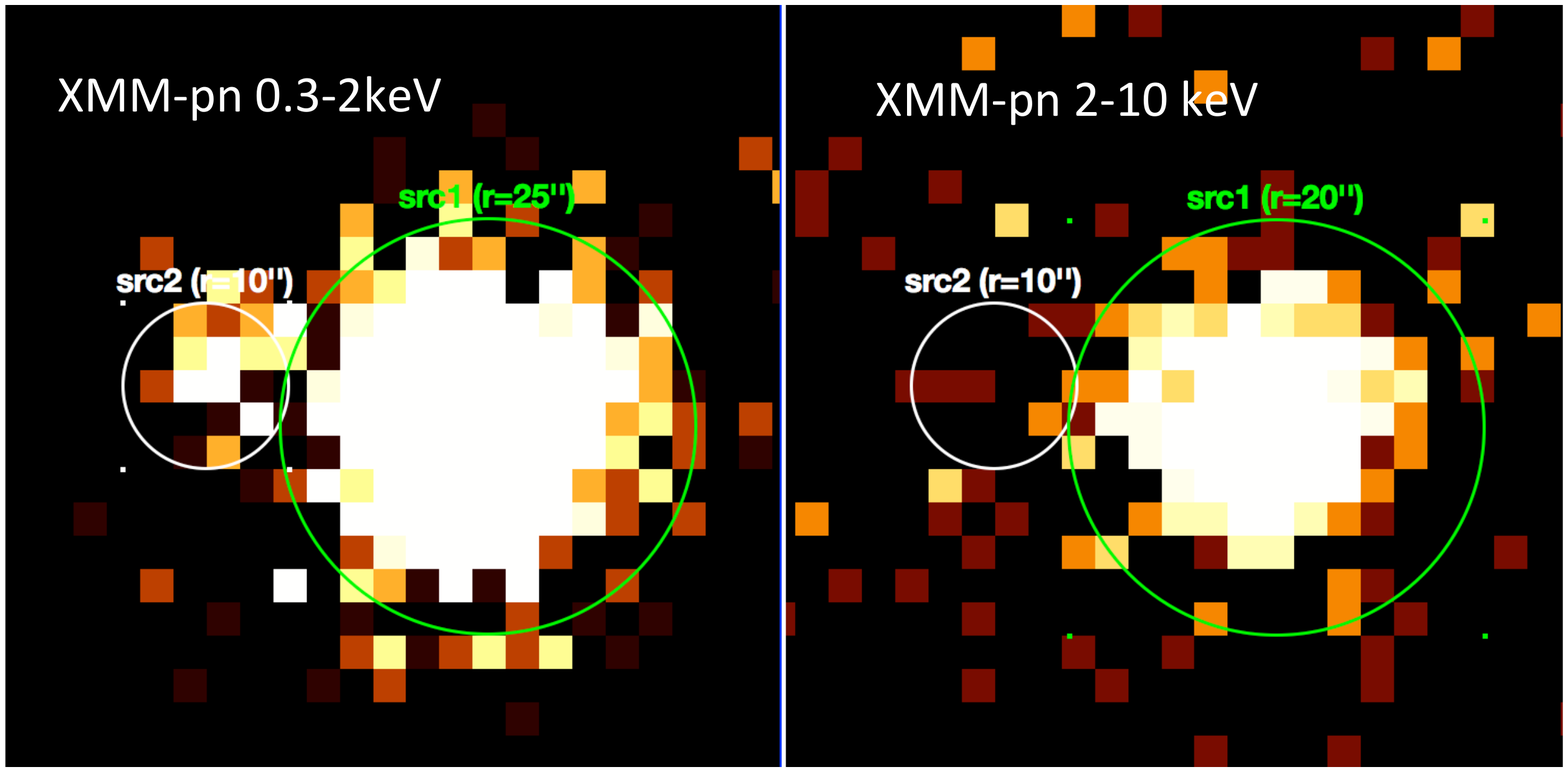}
\vspace*{-0.9cm}
{\footnotesize\caption{SDSS~J1038+3921. \textit{Left}: XMM-pn  map in 0.3-2 keV (left panel) and 2--10 keV (right panel).}}
\label{fig:1030_ima}
\end{figure*}

\begin{figure*}[]
\label{fig:spec1038}
\centering
\includegraphics[width=70mm,height=60mm]{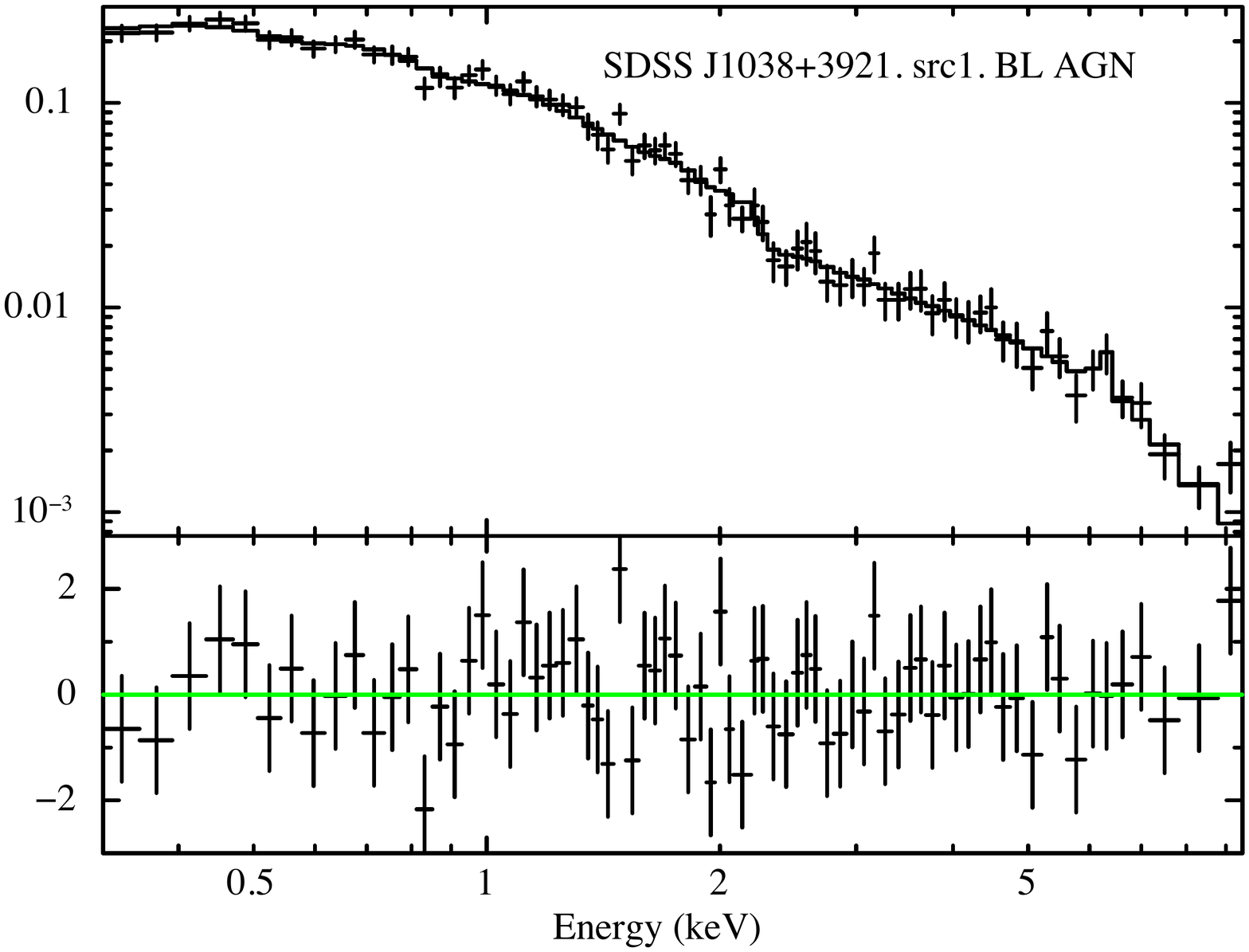}
\includegraphics[width=70mm,height=60mm]{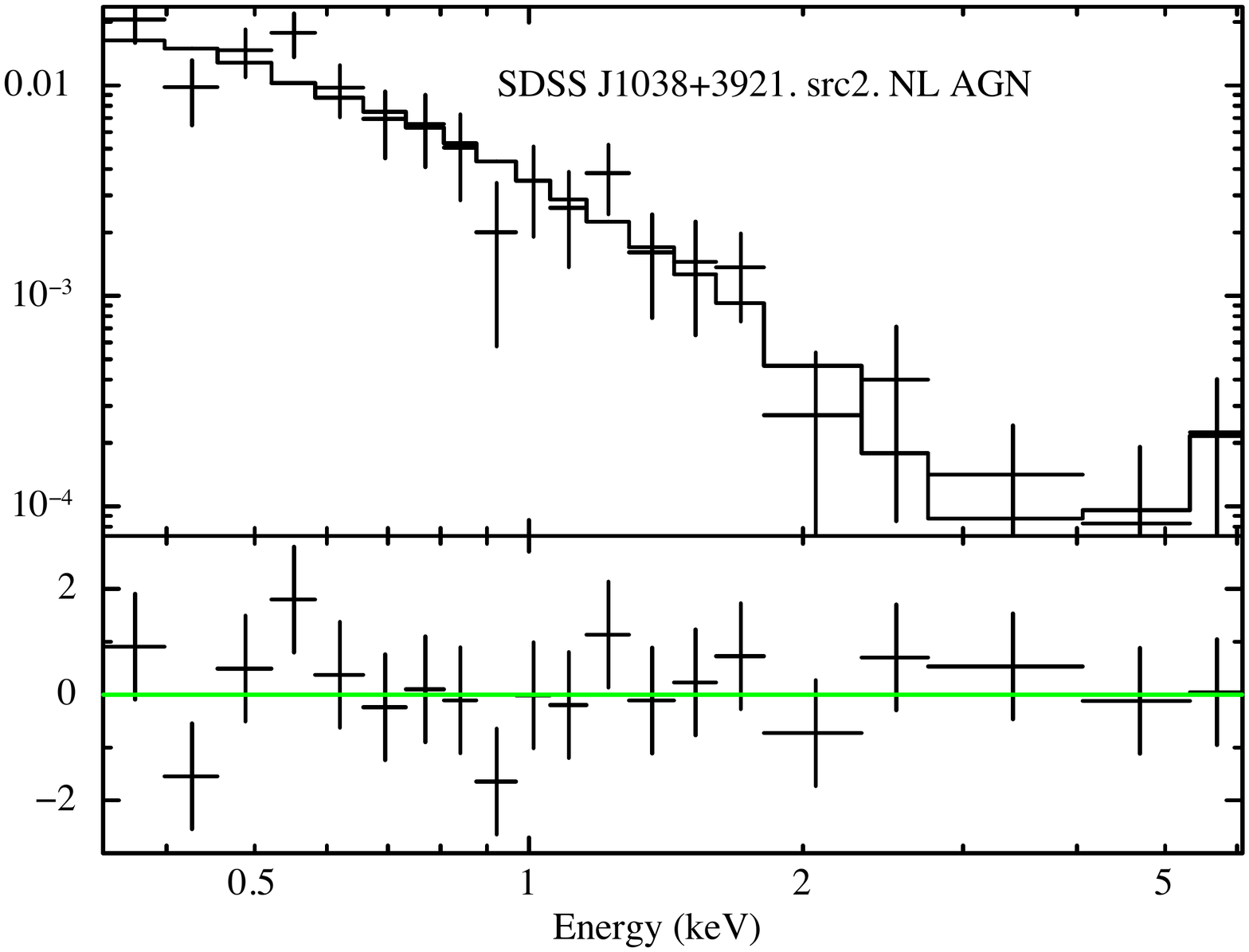}
\vspace*{-.4cm}
{\footnotesize\caption{SDSS~J1038+3921.  XMM-pn data of the sources detected in the system. Bottom panels show residuals of the best fit models and our data set in units of $\sigma$(see details in Sect. \ref{sect:j1038}). No strong residuals are present across the whole energy band.}}
 \end{figure*}

The main characteristics  of the system in SDSS~J1038+3921 are reported in Tab. \ref{tab:fluxes}.
From the optical spectrum, the objects in the system have been classified as a BL AGN (II-src1) and a NL AGN (II-src2). SDSS spectra for both sources are shown in Fig. \ref{fig:1038_opt}, while the X-ray image of SDSS~J1038+3921 is shown in Fig. \ref{fig:1030_ima}. 
The bright source (II-src1) clearly dominates the emission in the whole band, however the presence of II-src2 is evident mainly in the soft band (0.3--2 keV).
The X-ray spectrum of the BL AGN (II-src1) is shown in the left panel of Fig. \ref{fig:spec1038} and is reproduced with a primary power--law emission with cold and possibly warm absorber and  an intrinsic L(2--10 keV) =3$\times$10$^{42}$ erg s$^{-1}$. A Fe K line at 6.4 keV with equivalent width of about 200 eV is also detected. 
The X-ray spectrum of II-src2 is shown in the right panel of Fig. \ref{fig:spec1038} and is well reproduced with a highly absorbed power-law and a soft component emerging below 2 keV that is possibly due to photoionized emission  (Guainazzi \& Bianchi 2007). 
The observed 2--10 keV luminosity is about 10$^{41}$ erg s$^{-1}$. Due to the limited photon statistics in 2--10 keV and the XMM energy band limited to 10 keV, we are not able to constraint the absorption column density, then, in order to better characterize the nature of absorption in II-src2, we analysed archival multiband data and looked for absorption diagnostic. 
Fitting the optical spectrum of II-src2 we measure L([OIII])  corrected for extinction of 2$\times$10$^{41}$ erg s$^{-1}$, similar to the observed L(X). The evidence (L(X)/L([OIII])$\lesssim $ 1) may be strongly suggestive of a deep absorption  (see Cappi et al. 2006).
Assuming  a cold Compton-thick absorber (i.e. N$_H$=10$^{24}$ cm$^{-2}$) we obtain an intrinsic luminosity in (2--10 keV) of about 10$^{42}$ erg s$^{-1}$.
The fit of the optical-to-mid-IR spectral energy distribution (SED) of
II-src2 with a SED-decomposition code (including stellar
and AGN components; see Fritz et al. 2006; Feltre et al. 2012) provides a
12$\mu$m of $\approx2.8\times10^{42}$ erg s$^{-1}$ for the AGN component. This
converts into a predicted, intrinsic 2--10~keV luminosity of
$\approx2.5\times10^{42}$ erg s$^{-1}$ assuming the Gandhi et al. (2009) relation, in a very good agreement with L(X) we have measured in the case of deep absorption, strengthening the hypothesis of a Compton thick source (see e.g.  Lanzuisi et al. 2015).

From our multiband analysis  we  consider the more likely scenario of II-src2 as a bona-fide CT AGN in which the $>$10 keV emission is not detected. 
However, we cannot ignore another (even more intriguing) possibility of a  shut-off AGN, i.e. that  we are observing in the optical waveband the past activity of a now quiet BH.
It is worth noting that in all the X-ray reflection-dominated sources the only possible way to exclude this alternative hypothesis is either see the primary continuum piercing at hard X-rays (i.e. above 10 keV), or detect an unveiling event due to absorption variability. 
The fact that the other AGN is unabsorbed is also intriguing, somehow different with respect to most of the AGN pairs observed in X-rays (Bianchi et al. 2008, Koss et al. 2011, Guainazzi et al. 2005).

\section{Future prospects in the next XMM and Chandra decade} 
\label{sect:future}
We have presented our MAGNA programme that aims at studying the  multiband properties of a well selected sample of multiple AGN in different stages of merging.
We have discovered the CG \sdss\ characterized by exceptionally strong activity (five out eight galaxies are accretion powered).
Further studies of this CG are still on-going. At a mean redshift of z=0.0353, the NVSS and FIRST radio sensitivities already
allowed us to measure/constrain radio powers in the 10$^{21-22}$ W Hz$^{-1}$ regime. Our approved EVN observation  will go 
 more than an order of magnitude deeper in sensitivity allowing us to detect compact emission from the two LINERs (I-src2 and I-src3 in Tab. \ref{tab:fluxes}. 
Detection of these sources on 10-mas scales is expected because partially self-absorbed compact jets are ubiquitous in LINERs according to
VLBI studies, confirming that at least half of these are accretion-powered (Nagar et al. 2005);
in fact, the high X-ray luminosities of I-src2 and I-src3 are consistent with this prediction. 
The EVN and eMERLIN observations at 1.7 GHz will yield spatial resolutions of about 3 pc and 100 pc, respectively. Therefore, our observations will allow us to trace the pc-scale of the jet from $\sim$3-100 pc (eEVN), while eMERLIN will trace this emission above 100 pc scales. Our combined EVN-eMERLIN imaging will permit us to obtain an unprecedented view of the synchrotron emission powering each individual AGN in this remarkable compact group from the pc- to the kpc-scale.
We have also obtained integral-field spectroscopy data with MUSE on the VLT; these complement our high-resolution radio follow-up, as we will directly probe the impact of the jet on the kinematics of the ionized gas and compare the H$\alpha$-based star formation rates locally (jet-induced star formation) and globally. 
The deep IFU spectroscopy will allow us to map all galaxies and beyond to understand the cause for the enhanced activity of the group. In particular we will measure kinematic disturbances in the stellar and ionized gas kinematics, the stripped gas and bridges in relation to nuclear activity and star formation rate.\newline
It is worth noting that the high penetrative power of hard X-rays provides a unique and often ultimate tool in the hunt for multiple active nuclei in a galaxy, being less affected by contamination and absorption, and produced in large amounts only by AGN.
The high point like luminosity, the spectral shape of the continuum  and the presence of a strong Fe line emission in the X-ray spectrum are clear signatures of the presence of an AGN with respect to a region of starburst emission.
The clearest cases are the Chandra detected double nucleus in the luminous infrared galaxy NGC 6240 (Komossa et al. 2003) with a projected separation d$\sim$1 kpc, Mrk 463 (d$\sim$3.8 kpc, Bianchi et al. 2008), Mrk 739  (d$\sim$3.4 kpc, Koss et al. 2011); and the XMM-Newton detection of  IRAS 20210+1121 (d$\sim$11 kpc, Piconcelli et al. 2010),   ESO509-IG066 (d$\sim$10.5 kpc , Guainazzi et al. 2005), and possibly AM1211-465 (d$\sim$98 kpc, Jim\'enez-Bail\'on, 2007).
Some members of these systems show no (or very weak) explicit evidence of AGN  in their optical/near-infrared spectra. 
Together with SDSS J1038+3921, the XMM-OTAC  granted to our team in AO15 3 additional systems (4 systems with a total exposure of 180 ks).
The observation of the second system has been already performed and a preliminary analysis confirmed the presence of both nuclei. Detailed analysis are in progress.
The systems investigated so far and described in this paper have shown the effectiveness of our approach in order to study and characterize multiple AGN systems.
These results clearly show the importance of snapshot ($\sim$20-40 ks) observations with XMM and Chandra. The X-ray study potentially allows the detection and characterization of systems that are often elusive in the optical band. 
Thanks to its superior effective area up to 10 keV, XMM is particularly suited to detect the nuclei of systems at large separations, even in the case of strong absorption while the low background and sharp PSF of chandra will allow the study of systems in a close separation ($\lesssim$ 10 arcsec).
\acknowledgements
  All coauthors, members of the MAGNA team (www.issibern.ch/teams/agnactivity/Home.html) acknowledge support of ISSI-Bern, Switzerland.


\begin{thebibliography}{}
\bibitem[Baldwin et al. 1981]{Baldwin81}Baldwin et al. 1981, PASP, 93, 5
\bibitem[]{}Becker, R. H., White, R. L., Helfand, D. J. 1994, ASPC, 61, 165
\bibitem[Belgeman et al. 1980]{belgeman80} Begelman et al.1980, Nature 287, 307 
\bibitem[Bianchi et al. 2008]{bianchi08} Bianchi, S. et al. 2008, MNRAS, 386, 105,
\bibitem[Bogdanovic 2015]{bogdanovic15} Bogdanovic, T., 2015, ASSP, 40, 103
\bibitem[Bruzual \& Charlot 2003]{bruzual03}Bruzual G., Charlot S., 2003, MNRAS, 344, 1000
\bibitem[]{}Cappi, M., et al. 2006, A\&A, 446, 459
\bibitem[]{} Condon, J. J., wt al., 1998, AJ, 115, 1693
\bibitem[]{}Comerford, J., et al. 2012, ApJ, 753, 42
\bibitem[De Rosa et al. 2015]{derosa15} De Rosa, A., et al. 2015, MNRAS, 453, 214
\bibitem[Escala et al. 2004]{escale04} Escala et al. 2004, ApJ, 607, 765
\bibitem[Feltre et al. 2012]{feltre12} Feltre A, et al., 2012. MNRAS 426
\bibitem[Fritz et al. 2006]{fritz06 }Fritz J, et al., 2006. MNRAS 366:767–786 
\bibitem[]{}Fu, H., et al. 2011, ApJ, 733, 103
\bibitem[]{}Fu, H., et al. 2012, ApJ, 745, 67
\bibitem[]{}Gandhi, P. et al., 2009, A\&A 502, 457
\bibitem[Gonz\'alez-Mart\'inez et al. 2009]{gonzalez09} Gonz\'alez-Mart\'in, O, et al. 2009, A\&A, 506, 1107
\bibitem[Guainazzi et al. 2005]{guainazzi05}Guainazzi, M., et al 2005, A\&A 429, L9
\bibitem[]{} Guainazzi, M. \& Bianchi, S., 2007, MNRAS, 374, 1290
\bibitem[]{}Haehnelt, M.\& Kauffmann, G. 2002, MNRAS, 336, L61
\bibitem[Hickson 1997]{hickson97} Hickson, P. 1997, ARAA, 35, 357 
\bibitem[jimenez et al. 2007 1997]{jimenez07} Jim\'enez-Bail\'on, 2007, A\&A, 469, 881
\bibitem[]{}Nagar N.M.,et al., 2005, A\&A, 435, 521
\bibitem[Komossa \& Zensus 2016]{komossa16} Komossa, S. \& Zensus, J. A. 2016, IAUS, 312, 13
\bibitem[Komossa et al. 2003]{komossa03} Komossa, S., et al. 2003, ApJ, 582 L15
\bibitem[Koss et al. 2011]{koss11} Koss et al. 2011, ApJ 735 L42
\bibitem[]{} Lanzuisi, G., et al. 2015, A\&A, 578, 120
\bibitem[Liu et al. 2011]{liu11} Liu, X.; et al., 2011, ApJ, 737, 101
\bibitem[Piconcelli et al. 2010]{piconcelli10}Piconcelli, E.,  et al. 2010, ApJ, 722, L147
\bibitem[Ribeiro 1996]{ribeiro96} Ribeiro, A. L: B., De Carvalho, R. R., et al., 1996, ApJ, 463, L5.
\bibitem[]{} Rodriguez, C., et al., 2006, ApJ, 646, 49
\bibitem[]{}Satyapal, S., et al., 2014, MNRAS, 441, 1297
\bibitem[]{}Smith, K.L., et al. 2012, ApJ, 752, 63
\bibitem[Terashima \& Wilson 2003]{terashima03} Terashima \& Y., Wilson, A.S., 2003, ApJ, 583, 145
\bibitem[Turner et al. 2001]{turner01} Turner M. J. L. et al., 2001 A\&A, 365, L110
\bibitem[]{} Volonteri, M., et al. 2003, ApJ 582, 559

\end{thebibliography}
\end{document}